\documentclass[letter]{aa} 
\usepackage{natbib}
\usepackage{graphicx}
\usepackage{times}
\usepackage{amsmath}
\usepackage{txfonts}
\usepackage{graphicx}
\usepackage{wasysym}
\usepackage{txfonts}
\newcommand{\micron}{\,{\rm $\mu$m}}
\begin{document}
   \title{High-sensitivity search for clumps in the Vega Kuiper-belt
   \thanks{Based on observations carried out with the IRAM Plateau de Bure
Interferometer. IRAM is supported by INSU/CNRS (France), MPG (Germany)
and IGN (Spain).}}

   \subtitle{New PdBI 1.3\,mm observations
\thanks{Processed data from Fig.\ref{FigObs} are available in electronic form
at the CDS via anonymous ftp to cdsarc.u-strasbg.fr (130.79.128.5)
or via http://cdsweb.u-strasbg.fr/cgi-bin/qcat?J/A+A/}}
   \author{V. Pi\'etu
          \inst{1}
          \and
          E. Di~Folco\inst{2,3}
          \and
	  S. Guilloteau \inst{4,5}
	  \and
          F. Gueth\inst{1}
          \and
          P. Cox\inst{1}
          }

   \institute{IRAM, 300 rue de la piscine, F-38406 Saint Martin
              d'H\`eres, France \email{pietu@iram.fr}
              \and
              CEA Saclay/Service d'Astrophysique, Laboratoire AIM, CEA/DSM/IRFU-CNRS-Universit\'e Paris Diderot, F-91191 Gif-sur-Yvette Cedex, France
        \and
        LESIA-UMR 8109, CNRS and Observatoire de Paris-Meudon, 5 place J. Janssen, 92195 Meudon, France 
	\and
	Universit\'e de Bordeaux, Observatoire Aquitain des Sciences de l'Univers, 
		2 rue de l'Observatoire BP 89, F-33271 Floirac, France
  	\and
  	CNRS/INSU - UMR5804, Laboratoire d'Astrophysique de Bordeaux;  
		2 rue de l'Observatoire BP 89, F-33271 Floirac, France
              }

   \date{Received ; accepted }

 
  \abstract
   {Previous studies have found that Vega is surrounded by an extended 
   debris disc that is very smooth in the far infrared, but displays
   possible clumpiness at 850\,\micron\  and dust emission peaks at 1.3\,mm.}
   {We reobserved Vega at 1.3\,mm with PdBI to constrain its 
   circumstellar dust distribution.}
   {Our observations of a three-field mosaic have a factor of two higher 
    sensitivity than previous observations.}
   {We detect Vega photosphere with the expected flux, but none of the 
    previously reported emission peaks that should have been detected
    at the $>6 \sigma$ level, with a sensitivity $<$1\.mK.} 
   { This implies that the dust distribution around Vega is
       principally smooth and circularly symmetric. This also means that
       no planet is needed to account for dust trapped in mean-motion
       resonnance.}


   \maketitle
%

\section{Introduction}
More than 25 years after the first debris discs were discovered by the
IRAS mission \citep{aum84,aum85}, the presence of Edgeworth-Kuiper
belts around other stars has been widely attested. Statistical
studies conducted with the Spitzer satellite on large samples of main
sequence stars have shown that $30\pm5$\,\% of A type and $16\pm3$\,\%
of FGK type stars exhibit detectable IR excess emission longward of
24$\mu$m \citep{su06,bry06,tri08}. This excess luminosity is
attributed to the presence of cold dust grains arranged in a disc or
ring(s). Among those, only a handful of nearby systems have been
resolved because of obvious limitations in sensitivity and angular
resolution. Being one of the first systems detected, Vega (A0V,
7.76\,pc) has a particular status, and is usually considered as a
prototype of debris discs stars.

Given its proximity to the Earth, its face-on disc has been
successfully imaged from IR to mm-wavelengths. Its
appearance curiously seems to change from a smooth azimuthal profile at short
wavelengths to a more structured, clumpy ring in the sub-mm
domain. MIPS images at 24, 70 and 160$\mu$m \citep{su05} 
show circular profiles, the emission being detected up to 815\,AU for the latter. 
The radial profile analysis suggests a ring-like
distribution of material (after subtraction of the photosphere
contribution), with an inner boundary located around $85 \pm
15$\,AU. Herschel-PACS observations at 70 and 160$\mu$m
directly confirmed this ring morphology, with a peak intensity occuring 
at 85\,AU \citep{sib10}, as well as the smooth brightness profile.

   \begin{table*}[htb!]
   \begin{center}
   \begin{tabular}{l|c|c|c||c|c|c|c}
   \hline   
   Field      & RA Off.& Dec Off. & Noise& Fit. RA & Fit. Dec& Flux & Corr. Flux\\
     & (") & (") & (mJy) & (") & (") & (mJy) & (mJy) \\
   \hline
   \hline
   North-East    &  \phantom{-}3 & \phantom{-}5 & 0.26 &\phantom{-}0.13$\pm$ 0.16 & \phantom{-}0.30$\pm$0.14 & 1.70$\pm$0.23 & 2.05$\pm$0.28\\ 
   South-West & -3 & -5 &0.29 & -0.12$\pm$ 0.19 & \phantom{-}0.03$\pm$0.16 & 1.48$\pm$0.23 & 1.81$\pm$0.28\\
   Central & \phantom{-}0 & \phantom{-}0 & 0.37 &-0.70$\pm$ 0.15 &-0.01$\pm$0.16 & 1.96$\pm$0.32 & 1.98$\pm$0.32\\
   \end{tabular}
   \caption{Description of the fields observed and point-source fit. From left to right, the columns are: {\bf Col. 1:} Field name. {\bf Col.2,3:} Field pointing center (offsets from $\alpha=$18:36:56.330, $\delta=$38:47:01.30, J2000.0). {\bf Col. 4:} Noise in the field. {\bf Col. 5,6,7,8:} Fit of point-source. {\bf Col. 5,6:} Fitted positions are offsets from $\alpha=$18:36:56.359, $\delta=$38:47:01.20 (after correction for proper motion -- see text). {\bf Col. 7,8:} Fitted flux, and flux corrected for primary beam attenuation.
   \label{tab:flux}}
   \end{center}
   \end{table*}

At sub-mm wavelengths, the first image was reported by \citet{hol98}
with JCMT/SCUBA at 850$\mu$m. The resolved emission extents to scales
twice as large as our Solar System, with a possible peak emission
located at $\sim70$\,AU, northeast of the star, accounting for less
than 40\,\% of the total flux. The possible asymmetry of their map,
although at the 2\,$\sigma$ level, has motivated several subsequent
studies to explore the nature of the blob and its putative link with
unseen planets. \citet{Koerner_2001} observed Vega with OVRO at
1.3\,mm and reported a ring arc located at 95\,AU (12\arcsec from the
star).  \citet{Wilner_2002} used the Plateau de Bure Interferometer
(IRAM) to probe the continuum emission at 1.3 and 3.3\,mm. While the
star was only detected at the former wavelength, no peak emission
appeared in the original resolution map. Degrading the interferometer
resolution, \citet{Wilner_2002} claimed that extended exists
emission concentrated in two blobs located at $8$ and $9.5$\arcsec (60
and 75\,AU), southwest and northeast of the star respectively. Using
the SHARC~II camera at CSO, \citet{mar06} detected ring-like
structures of the 350$\mu$m and 450$\mu$m emissions, with some
inhomogeneity at the 2-4\,$\sigma$ level.  The brightness peaks in the
photosphere-subtracted images partially overlap with those reported at
850$\mu$m and 1.3\,mm, strongly suggesting that grains
and/or planetesimal concentrations exist in the belt around Vega.

The possible existence of concentrations in the dust belt has long
been interpreted (and modelled) as the evidence of grains trapped in
mean-motion resonances (MMR) with a putative, but unseen, massive
planet located close to the inner radius of the planetesimal ring
\citep{hol98,Wilner_2002,wya03}.  Two scenarios have been invoked to
populate the MMRs with an unseen planet: the particles are trapped
either while migrating inward under the influence of
Poyinting-Robertson (P-R) drag \citep[e.g.][]{kuc03}, or during the
outward migration of the planet itself.  Models developed by
\citet{wya03} and \citet{rec08} invoke the outward migration of a
Neptune- to Saturn-mass planet in a circular (and elliptical, respectively)
orbit to reproduce the observed clumps.  To reconcile the
different views of the disc morphology at the different wavelengths,
it has been suggested that the clumps seen in the sub-mm trace large
grains trapped in the same resonance as their parent planetesimals,
while intermediate and small grains (down to the blown-out size) are
decoupled from the resonant structure, thus forming a smooth and
axisymmetric disc component observed at shorter wavelengths. This
broad and smooth halo could therefore be sculpted by small grains
evacuated from the MMRs by the stellar radiation pressure
\citep{wya06} and/or by the high-velocity collisions between their
parent planetesimals \citep{kri07}.  Substantial observational efforts
have been invested in the quest for the predicted planet(s), mostly
based on near-IR adaptive optics (and/or coronagraphic) techniques on
large telescopes \citep{maro06,hin06}. This search remains
unsuccessful today, essentially because of the lack of sensitivity at
the predicted planet position ($7-10$\arcsec).

The Vega system thus appears to be a unique laboratory for 
constraining and understanding the evolution of dust grains in exo-Kuiper
belts, in particular the interaction between planetesimals, grains, 
and planets.
 

\section{Observations and analysis}

\subsection{Observations}

During the winter 2006/07, a new generation of receivers was installed
on the Plateau de Bure interferometer, enabling single frequency
dual-polarization observations with improved noise figures,
single-side band tuning and larger bandwidth. This instrumental
upgrade resulted in a significantly improved sensitivity of the array
(at 1.3\,mm, the total gain was an increase by a factor of four in
continuum sensitivity). As part of the science verification program,
Vega was observed in January 2007 at 230.538 GHz.
These observations were carried in the D (compact) configuration of
the PdBI (with baselines ranging from 15 to 85\,m, the most suited arrangement 
for detecting large scale emission). The tuning was in the lower side-band
(LSB), with $T_{rec} \sim 50$~K and a bandwith of 850 MHz in each of
the two orthogonal linear polarizations (H and V).  The atmospheric
phase noise ranged from 20 to 50$^{\circ}$.  Despite these first observations 
reaching a sensitivity comparable to that of \citet{Wilner_2002}, 
we detected only the central point source, but no
extended emission at all.

We thus reobserved a two-field mosaic towards Vega (offset by (3 ,
5\arcsec) and (-3 , -5\arcsec), respectively, the primary beams being indicated in
Fig.\ref{FigObs} together with the central pointing obtained in
January) in October and November 2007, again in the D configuration
and with the same spectral setup. Atmospheric phase noise was again
between 20$^{\circ}$ and 50$^{\circ}$. The absolute flux calibration was
performed using observation of MWC349. Its flux density model has
recently been revised at IRAM, with a 16\% increase in flux (Krips et al. in
prep), so we used 1.98\,Jy as a flux model at 230.5\,GHz. With a
grand-total observing time of 16 hours on source, we finally reach a
sensitivity limit below 0.2\,mJy/beam (0.9\,mK with natural weighting)
in the central region of the final mosaic map, and $<0.4$\,mJy/beam
within 11\arcsec\ from Vega.

\subsection{Analysis}

Vega has a high proper motion ($\mu_{\alpha} = 201.03$~mas/yr,
$\mu_{\delta} = 287.47$\,mas/yr, with an error ellipse of
$0.63\times0.54$\,mas/yr, at PA 144$^\circ$, \citealp{Perryman_1997}). 
We corrected the individual ({\it u,v}) tables for proper motion using
the value of \citet{Perryman_1997} to a common J2000.0 Equinox prior
to data analysis.

We then performed point-source fits in the ({\it u,v}) plane and detected a
point source in each of the fields. The fitted position agrees within
the error bars (see Table \ref{tab:flux}), as does the flux density, 
when we take into account the attenuation by the primary beam
for the offset fields. The mean value is $1.95\pm0.17$\,mJy.
\citet{Wilner_2002} found 1.7$\pm$0.30\,mJy
\footnote{They cite 1.7 $\pm$ 0.13 mJy from a Gaussian fit, but the
  error in a point source flux cannot be below the rms noise of the
  image, which they indicate is 0.3 mJy/beam}, but used 1.70\,Jy as
a flux density for MWC349. Correcting for the calibrator flux
difference, their flux becomes 1.98$\pm$0.35\,mJy, in excellent
agreement with our measurement.  This is consistent with the 2.11\,mJy 
value expected from the pole-on photosphere for which we refer 
to the modelling by \citet{muel10}, who took into account the discrepancy 
between the polar and equatorial temperature resulting from the fast
spin velocity of this star.

   \begin{figure*}
   \centering \includegraphics[width=18cm]{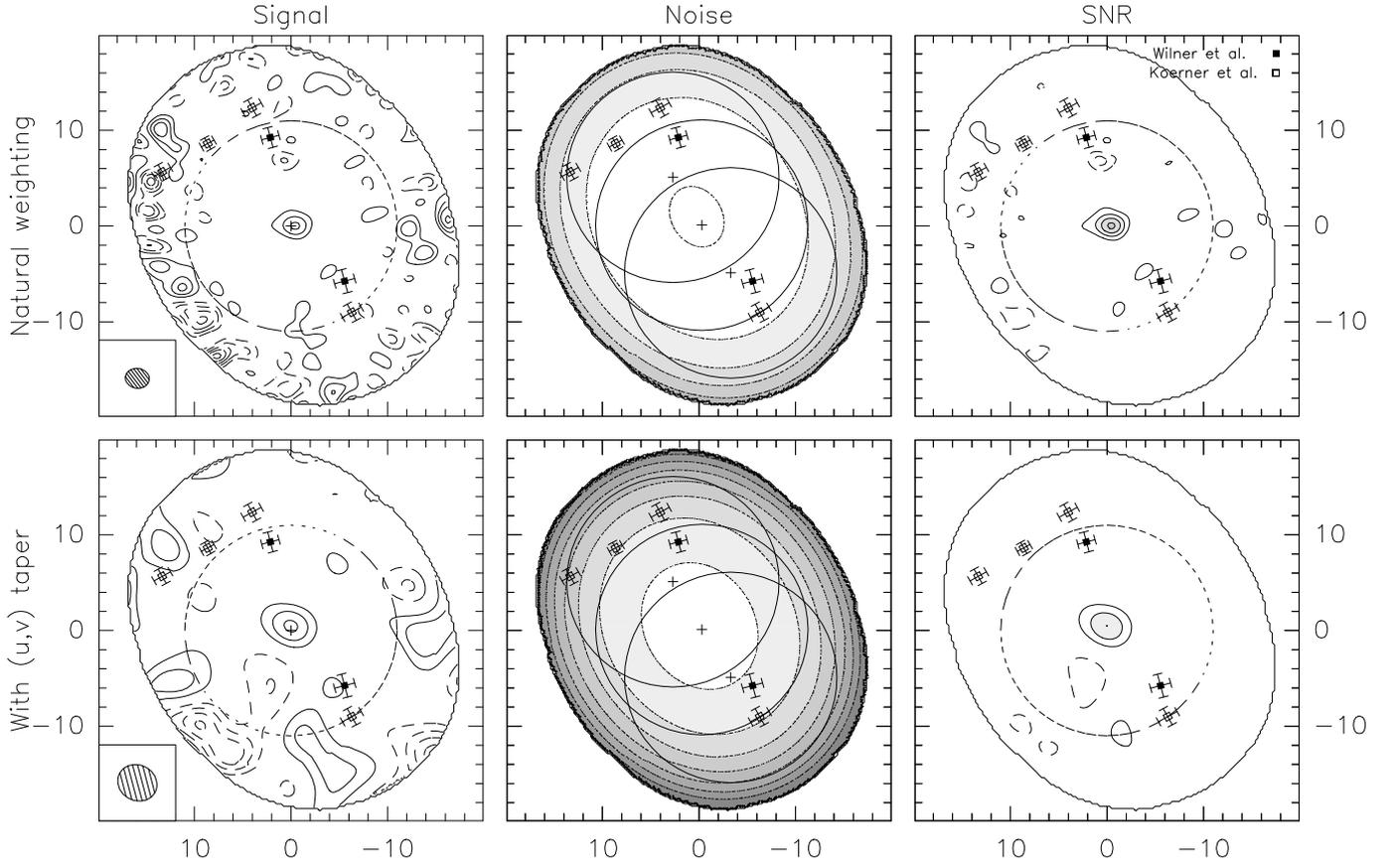}
   \caption{{\bf Top left:} natural weighting 1.3\,mm continuum image
     (corrected for the primary beam attenuation). The angular
     resoluion is 2.5$\times$2.1'' at PA 82$^\circ$ and the contour
     spacing 0.5\,mJy/beam (2.1\,mK), with negative contours being
     dashed.  The dashed circle represents the primary-beam half-power
     field of view. {\bf Top center:} corresponding noise map. Contour
     spacing is 0.2\,mJy/beam . The circles show the observed
     individual fields (half-power field of view, centered on the
     crosses). {\bf Top right:} signal-to-noise ratio map. Contour spacing
     is 2\,$\sigma$. {\bf Bottom left:} 1.3\,mm continuum image
     obtained with a ($u,v$) taper. Angular resolution is
     4.2$\times$3.7'' at PA 60$^{\circ}$ and contour spacing is
     0.6\,mJy/beam (0.8\,mK). {\bf Bottom center:} corresponding noise
     map. Contour spacing is\,0.2 mJy/beam (with the first contour
     being at 0.4\,mJy/beam). {\bf Bottom right:} signal-to-noise ratio
     map. Contour spacing is 2\,$\sigma$. In all maps, the filled
       square symbols indicate the location of the extended emissions
       found by \citet{Wilner_2002} and the empty square symbols
       indicate the location of the peaks reported by
       \citet{Koerner_2001}.  North is up, and east to the left. 
\label{FigObs}}
    \end{figure*}

Data were processed with GILDAS software\footnote{See
  \texttt{http://www.iram.fr/IRAMFR/GILDAS} for more information about
  GILDAS.} using the classical mosaicing algorithm \citep[see
  e.g.][]{Gueth_2000}.  We present the resulting continuum maps in
Fig.~\ref{FigObs}: in the top row we present maps obtained with
natural weighting, while in the bottom row the data have been tapered in
the ({\it u,v}) plane (with elliptical Gaussian $1/e$ widths of
$30\times40$\,m at PA 0$^{\circ}$) to increase the
sensitivity to putative extended emission by degrading the angular
resolution.  Since the mosaicing algorithm produces images corrected
for the primary beam attenuation, the noise increases towards the
map's edges (as is clearly visible in the signal maps in the left column of
Fig.~\ref{FigObs}). We present in the middle column of
Fig.~\ref{FigObs} the corresponding noise maps that show our
sensitivity limit as a function of the position in the field of
view. The right column of Fig.~\ref{FigObs} shows the signal-to-noise
ratio maps obtained in each case by dividing of the signal map by the
noise map. From these, we conclude that only the central point source
is detected with a SNR$>3$.

\subsection{Comparison with previous work}
\label{sec:comp}
 Previous interferometric observations at 1.3\,mm were reported by
  \citet{Koerner_2001} and \citet{Wilner_2002}.  With an angular
  resolution of 3.3 x 2.9\arcsec, \citet{Koerner_2001} found an arc
  ring at 95 AU with four peaks located between 11 and
  14.5\arcsec\ of flux density ranging from 2.4 to 4.0\,mJy (3 to
  4$\sigma$ level) after correcting for primary beam attenuation.  In
  addition to the central point source, \citet{Wilner_2002} reported
  two emission peaks located at 8.0 and 9.5\arcsec, after tapering
  their data to a 5.3 x 4.6\arcsec\ resolution. From a Gaussian fit in the
  {\it (u,v)} plane, they obtained fluxes (corrected for the primary
  beam attenuation) of $7.1\pm1.4$ and $4.3\pm1.0$\,mJy, respectively
  (4 to 5$\sigma$ level). The position of these dust peaks is reported
  in Fig. \ref{FigObs}, with their $1 \sigma$ error ellipse derived
  from the signal-to-noise ratio and synthesized beam.  These two results clearly
  do not agree, as none of the error ellipse overlap.  We also note that
  the comparison of the different results need to take into account
  that their single field maps are not corrected for primary
  beam attenuation, in contrast to the maps we present here.

At the position of the peaks reported by \citet{Koerner_2001}, our
observations have a 1$\sigma$ sensitivity of 0.35 mJy/beam (0.54 for
the most distant peak), a factor of between two and three better than the OVRO
observations. If real, these peaks should have been detected at the $>
6-8\sigma$ level. Similarly, our observations reach a $1\sigma$
sensitivity limit of 0.26 mJy/beam (1.1 mK) or 0.45mJy/beam (0.6 mK)
at the position of the \citet{Wilner_2002} peaks for the untapered and
tapered maps, respectively.  Thus, depending on their size, the two
structures should have been detected at the $10-15 \sigma$ and $6-10
\sigma$ level in our maps. 


Among the possible ways of explaining our non-detection, we exclude
calibration errors since we detect the central source, at the expected
position and with the expected flux, which is an excellent internal
quality check.  Given the above-mentioned flux calibration scaling
with respect to the \citet{Wilner_2002} data, any supposed bias in the flux
calibrationshould enhance the probability of detecting the blobs in our maps,
since their intensity should also be higher by 16\,\%.  If these structures 
had rotated according to Keplerian motion, the displacement within 
the $\sim$~6 year interval would have been smaller than 1\arcsec.

\section{Discussion}
Our PdBI observations are consistent with no detectable millimetric
emission from the outer dusty disc (or ring) surrounding Vega. 
We  briefly discuss the impact of this new constraint on the dust
properties in this system.

\subsection{Clumpiness}
We have shown in section \ref{sec:comp} that the clumps previously
reported at 1.3\,mm by \citet{Koerner_2001} and \citet{Wilner_2002}
are not detected in our observations, which are deeper in sensitivity by a
factor of two, and are probably low-significance artifacts. A trivial
implication of this finding is that no planet is required to trap dust
in mean-motion resonance at these positions.  More generally, our
results also shed new light on the clumpiness of the dust surrounding
Vega. Observations at shorter wavelengths display very smooth images
\citep{su05,sib10} and the asymmetry of the 850\micron\ image is only
at the 2\,$\sigma$ level, hence also compatible with a smooth dust
distribution \citep{hol98}. The only detection of a possible asymmetry
comes from the SHARC~II images at 350 and 450\micron\ \citep{mar06}
but is at the 2-4\,$\sigma$ and $<10$\% of the ring brightness. In
conclusion, all observations display or are compatible with a smooth
dust distribution.

\subsection {Upper limit on the disc emission at 1.3\,mm}
Earlier sub-mm observations have convincingly established the
existence of an extended emission around Vega, with a ring-like
morphology at 350\micron, 450\micron\ with CSO SHARC~II \citep{mar06},
and 70\micron-160\micron\ with Herschel/PACS \citep{sib10}. According
to these studies, the peak intensity is located around $85-100$\,AU
($11-13$\arcsec). To set an upper limit on the 1.3\,mm
emission of this ring, we fitted a thin, uniform, 85\,AU radius ring
into the {\it (u,v)} data, carefully taking into account the different
fields of the mosaic (for numerical reasons, this thin ring is
represented by an adequately sampled series of point sources). 
  We find an integrated flux of $0.1\pm2.0$\,mJy, i.e. a 3 $\sigma$
  upper limit of 6 mJy. The lack of clear detection at high
  angular resolution relative to the bolometer results, which are more
  sensitive to extended flux, suggests that a thin ring may not be the
  most appropriate distribution.  If we instead assume a uniform
  annulus of inner radius 85\,AU and outer radius 105\,AU, we find an
  integrated flux of $-5.3\pm2.4$\,mJy, i.e., a $3 \sigma$ upper limit
  of 7.2 mJy, which is similar to the narrow ring value.

 Using the \citet{sib10} model, with an inner Gaussian and outer
  exponential decline in surface brightness (with a transition
  radius of 14\arcsec), our results are compatible with a total flux
  $< 24$ mJy (3 $\sigma$) if the distribution is truncated at 200 AU,
  and $ <30 $mJy if it extends up to 300 AU. This modelling also 
  agrees with the $3\sigma$ upper limit of 6\,mJy for emission
  within 105 AU. With this model, we are clearly unable to place 
  strong constraints on the dust properties, since our data lack
  sensitivity to faint extended emission.

Assuming a dust temperature of 68\,K \citep{Sheret_2004} and a dust
absorption coefficient of 0.5\,cm$^2\cdot$g$^{-1}$ \citep{Natta_2004}, 
our 3\,$\sigma$ flux density limit of 6\,mJy translates into an upper
limit on the dust grain mass in the ring of $M_{\rm dust} = 0.8
M_{\textrm{moon}}$ (or $10^{-2} M_{\bigoplus}$, not taking into
account the parent bodies). With the \citet{sib10} model, the upper
limit on the total dust grain mass is 3.9 $M_{\textrm{moon}}$. This
upper limit is consistent with the maximum dust mass obtained by
\citet{muel10} in their modelling of the Vega disc as a steady-state
collisional cascade.

 The best fit of the dust spectral energy distribution using
  measurements of \citet{sib10} and \citet{hol98}, corrected for the
  stellar emission, gives a spectral index of 2.9 and results in an
  extrapolated flux at 1.3\,mm of 12\,mJy. Our upper limit hence 
  agrees with the measurements made at shorter wavelengths.

\section{Conclusions}
Taking advantage of new receivers at IRAM Plateau de Bure
Interferometer, we have probed the Vega debris disc to investigate in
depth the possible presence of clumps in the well-known dust
belt. Apart from the stellar photosphere, we have detected neither
a point-source nor extended emission in our three-field mosaic, although our
sensitivity limit at 1.3\,mm has improved by a factor of two
compared to earlier studies. {\bf In particular, 
we have found no evidence for the northeast and southwest blobs 
claimed by \citet{Koerner_2001} or \citet{Wilner_2002}. }
These structures, which had been interpreted as the signature of the 
gravitational influence of putative unseen planets, should have been 
detected in our interferometric map at the $>6\sigma$ level.

The upper limits that we have derived on both the emission of a narrow ring
  extended emission are compatible with observations at
  shorter wavelenghts. All observations indicate that the disc/ring
  system surrounding Vega is smooth and circularly symetric, hence
  do not require the gravitational influence of a giant planet to
  sculpt asymmetries in the disc, at the current level of detection.

\begin{acknowledgements}
We acknowledge help from the Plateau de Bure staff when carrying out
the observations. We thank our anonymous referee for helpful comments.
\end{acknowledgements}

\bibliography{vega}
\bibliographystyle{aa}
\end{document}